# Nanocluster ionization energies and work function of aluminum, and their temperature dependence

Avik Halder and Vitaly V. Kresin

*Department of Physics and Astronomy, University of Southern California,
Los Angeles, California 90089-0484, USA*

**ABSTRACT**

Ionization threshold energies of $Al_n$ ($n$=32-95) nanoclusters are determined by laser ionization of free neutral metal clusters thermalized to several temperatures in the range from 65 K to 230 K. The photoion yield curves of cold clusters follow a quadratic energy dependence above threshold, in agreement with the Fowler law of surface photoemission. Accurate data collection and analysis procedures make it possible to resolve very small (few parts in a thousand) temperature-induced shifts in the ionization energies. Extrapolation of the data to the bulk limit enables a determination of the thermal shift of the polycrystalline metal work function, found to be in excellent agreement with theoretical prediction based on the influence of thermal expansion. Small clusters display somewhat larger thermal shifts, reflecting their greater susceptibility to thermal expansion. Ionization studies of free size-resolved nanoclusters facilitate understanding of the interplay of surface, electronic, and lattice properties under contamination-free conditions.

# I. INTRODUCTION

The energy required to remove an electron from an individual atom, molecule or atomic cluster (the ionization energy) or from a solid (the work function) is one of the main characteristics of the physical system. Metal clusters share with atoms the existence of a set of discrete electronic orbitals (in particular, so-called "superatom" clusters which display a clear electronic shell structure pattern [1,2]), and they share with solids the presence of an ionic framework which supplies a bath of thermal excitations. As a result, cluster ionization energies, like solid surface work functions, become temperature dependent. While the inherent thermal work function shift is usually not strong (as opposed to changes occurring when temperature-induced surface reactions or migrations take place [3]), it is an interesting quantity because it derives from an interplay between structural and electronic degrees of freedom. This type of interplay is an important contributor to a variety of physical processes.

Theoretical models (see [4] and references therein) accounting for the effects of thermal expansion have made specific predictions about the magnitude of the inherent effect in various metals. Accurate measurements of work functions as a function of temperature are therefore quite instructive. At the same time, such measurements are challenging because they impose very strict requirements on the quality and purity of the surface. Indeed, even a minute amount of contamination, which itself is temperature-dependent, can mask the effect completely [5]. The situation is especially challenging with the surfaces of reactive metals.

In the above contexts, aluminum and its nanoclusters, the subject of this paper, are of unquestionable relevance. On one hand Al is often viewed as a nearly-free-electron trivalent metal, but on the other hand its lattice potential modifies the shape of the Fermi surface in the bulk [6] and introduces notable perturbations to the free-electron-like shell structure ordering and response behavior in the small cluster limit (see, e.g., [7,8]). At the same time, the aluminum surface is highly prone to oxidation, hence the ability of the nanocluster beam technique to interrogate the particle on a short enough time scale to avoid contamination is very advantageous. Here we report on the first systematic size-resolved study of the temperature dependence of ionization threshold energies of metal clusters, and apply it to aluminum.

The plan of the paper is as follows. In Sec. II we briefly outline the measurement approach and describe the procedure for extracting accurate ionization energy values from the photoionization yield curves. In Sec. III we present the size-dependent aluminum cluster ionization energies and discuss their size and temperature variation, as well as the deduced thermal shift of the extrapolated bulk metal work function. Sec. IV summarizes the results and comments on potential future advances.

# II. EXPERIMENTAL METHOD AND DATA ANALYSIS

## A. Measurements

We briefly outline the experimental approach, which has been described in detail previously [9-11]. Neutral $Al_n$ clusters are produced by a magnetron sputtering/gas condensation source [12,13]. An important addition is a specially designed thermalizing tube at the end of the cluster aggregation chamber. The walls of the tube can be set to a well-controlled temperature in the range from 65 K to 215 K, and then the clusters equilibrate very close to the wall temperature thanks to collisions with the gas molecules during transit through the tube. In this experiment, temperatures of 65K, 90K, 120K, 170K and 230K were used.

Clusters exiting the thermalizing tube pass through a skimmer into the extraction region of a linear Wiley-McLaren time-of-flight mass spectrometer where they are ionized by 5 ns pulses from a tunable Nd-YAG/OPO laser system. The ionization wavelength is varied from 210-250 nm in a random sequence (to avoid any systematic bias). Laser pulse energy is carefully monitored immediately past the ionization region, and the ion yield curves are corrected for the intensity drifts of both light and cluster beams. A fluence of ≈500 µJ/cm$^2$ ensures single photon absorption, as verified by the linear dependence of the ionization yield on the laser pulse energy. Extracted cluster ion packets are detected by a channeltron equipped with a high-voltage conversion dynode, recorded by a multichannel scaler, and fitted by Gaussian peak deconvolution to form a mass spectrum.

By following the intensity of a given cluster's mass peak as a function of photon energy $E$ we can map out, for a preset cluster temperature, its photoionization yield curve $Y_n(E)$. Resolving the very small thermal shifts requires paying attention to the stability and accuracy of the data collection, hence each yield curve measurement for a given temperature lasted approximately 24 hours and was repeated 4-5 times. An example is shown in Fig. 1(a). Since we are working in the near-threshold one-photon regime, ionization-induced fragmentation should not be a concern, and we can assign the appearance thresholds to vertical ionization energies of the corresponding neutral clusters.

## B. Threshold analysis

Deriving ionization energies $I_n$ from ionization yield plots is not elementary because of the lack of a general theory of cluster photoionization. A number of more or less *ad hoc* fit algorithms can be found in the literature (see, e.g., [1,14-17] and references therein), a common one being simple linear extrapolation of $Y(E)$. On the other hand, an excellent description of nanoparticle work functions is provided by the Fowler law of photoemission which was originally developed for metal surfaces [5,18], then applied to aerosol-scale particles [19-22], and more recently to smaller metal clusters [9,23,24]. It must be added, however, that Fowler's original derivation was based on integrating the photoelectron flux over a continuous density of states, and a rigorous justification for applying it to a cluster with discrete shell structure is an interesting theoretical problem.

Since the temperature of the aluminum clusters in the present study is low, the Fowler formula can be approximated by the quadratic rise

$$Y_n(E) \propto (E-I_n)^2. \qquad (1)$$

The ionization energy can therefore be determined from a straight line intercept for $[Y_n(E)]^{1/2}$ near threshold. Such a procedure was used previously for the determination of work functions of cold alkali nanoparticles [25] and copper clusters [9] in a beam. Fig. 1(b) illustrates that a very good match to the data is indeed obtained.

Since we are interested in resolving very small shifts in the ionization thresholds (on the order of few tenths of one percent, see below), a careful uniform fitting procedure is required because even a dose of irregularity can wash out the effect of interest. For example, it is important to ensure that the interval over which the fitting is performed remains unchanged for all temperatures for a given cluster and that it is chosen so as to provide a stable fitted $I_n$ value.

To ensure the requisite stability, the following procedure was followed. Individual $Y_n(E)$ curves from different runs for a specific cluster size and temperature were interpolated in 10 meV segments by cubic spline fitting and averaged. The interpolation was useful because runs differed slightly in the number of included wavelength points, and it was important to achieve maximum uniformity in the data array being fitted. Having done this, the square root of the yield curve was plotted and a near-threshold energy interval $\Delta E_n$ was chosen for each cluster size in such a way that the plot of $[Y_n(E)]^{1/2}$ would be linear within $\Delta E_n$ for all temperatures. The intervals varied from cluster to cluster, and their average span was ≈0.2 eV. As mentioned in the previous paragraph, these energy intervals were then held fixed for the linear fitting procedure. At this point, in order to guarantee that the deduced $I_n$ is not excessively skewed by some individual data points, $\Delta E_n/(10\ \text{meV})$ points were randomly picked from within the selected energy interval and the $x$-intercept of the corresponding straight line fit determined. By rerunning such a selection 100 times and averaging the results, we found conclusive convergence to an accurate value of $I_n$.

The next section discusses the resulting set of size- and temperature-dependent values of ionization energies $I_n$.

### III. RESULTS AND DISCUSSION

#### A. Ionization energies

In Fig. 2 we first plot the ionization energies of $Al_{n=32-95}$ clusters averaged over the set of five surveyed temperatures.

Oscillations in the $I_n$ values reflect variations in the stabilities of the clusters' electronic configurations. Drops following $n=46,56,66,78$ are fully consistent with the shell model's spherical shell closings for 138,168,198,234 electrons [26], while those at $n=36$ and 38 are commensurate with shell closings at 106 and 112 electrons when one considers that neutral aluminum clusters can contribute electrons only in multiples of 3. However, not every possible shell configuration is reflected in the ionization energy plot, and conversely, some sizeable drops such as at $n=42$ and 54 have no obvious spherical shell counterparts. As mentioned in the Introduction, this appears to reflect the role of the ionic core in $Al_n$ clusters, where it can split and/or hybridize the electronic shells.

The pattern of ionization energies on the whole matches that observed in previous work [27-29], also marked in the figure, and the magnitudes also agree to within several percent or better. It should be reiterated that the $I_n$ values in the present work derive from highly systematic data collection and analysis procedures, and it is important that they extrapolate very closely to the standard polycrystalline work function value (see below).

#### B. Thermal shifts

The thermal variation observed for most clusters is ~0.2%-0.5% of the ionization energy. Data for each size at each investigated temperature are provided in the Supplemental Material [30]. Fig. 3 shows how $I_n$ values change from the minimum (65 K) to the maximum (230 K) of the investigated temperature range (note the change in the size scale compared to Fig. 2).

Generally speaking, the smaller clusters exhibit a somewhat larger temperature shift than the larger ones. Since inherent work function variations appear to be due largely to the effect of thermal expansion (which reduces the effective electron density and the Fermi energy [4]), this trend makes sense: the greater surface-to-volume ratio of the smaller clusters imparts to them a larger thermal expansion coefficient [31]. Analogously, it was reported recently that the compressibility of small noble-gas clusters appears to exceed the bulk value [32].

The cluster ionization energies can be extrapolated to the bulk limit ($n^{-1/3} \to 0$), recovering the metal surface work function $\varphi$. The lines in Fig. 3 demonstrate that the commonly used (see the reviews [1,33,34]) size dependence

$$I_n = \varphi + \alpha e^2/R, \qquad (2)$$

provides a good fit to the average trend. Here $R$ is the cluster radius, parametrized as $R = r_s a_0 (3n)^{1/3}$, where $a_0$ is the Bohr radius, $r_s$ is the electron density parameter (for aluminum [6] $r_s = 2.07$), and $3n$ is the number of valence electrons in the cluster. We find that the slope of the fitted lines is $\alpha = 0.34$, which is quite close to the frequently cited theoretical value of $\alpha \approx 3/8$ based on a semiclassical amalgamation of the bulk work function barrier and the image potential of a spherical metal particle (see, e.g., [34]).

The extrapolated values of $\varphi$ are 4.349 eV at 65 K and 4.338 eV at 230 K, in very good agreement with the polycrystalline room-temperature value of 4.28 eV quoted in the literature [35]. The corresponding drop in the work function over this experimental 165 K temperature difference is $\Delta\varphi = -10.6$ meV.

According to the theoretical analysis in Ref. [4], polycrystalline work function shifts between two temperatures can be represented by $\Delta\varphi = A\Delta T + B\Delta(T^2)$. Using their suggested $A$ and $B$ coefficients for aluminum ($-4.4 \times 10^{-5}$ eV/K and $-6.5 \times 10^{-8}$ eV/K$^2$, respectively), the predicted magnitude of $\Delta\varphi$ from 65 K to 230 K is $-10.5$ meV. The agreement with our measurement is excellent.

We can also quantify the variation of the clusters' ionization energy with temperature. Within the accuracy of the measurement, a linear fit to the data points shown in the Supplementary Material [30] is sufficient. The average over all cluster sizes in the range $n=32-95$ is found to be $\langle dI/dT \rangle_n \approx -7.8 \cdot 10^{-5}$ eV/K, the minus size denoting the fact that the cluster ionization energies also predominantly decrease with temperature (cf. Fig. 3). Note that whereas cluster-to-cluster variations in the data may need to be viewed with care, the average shift over the studied size range can be considered quite reliable.

This slope, which incorporates also the smaller cluster sizes in the investigated range, is larger than the effective one for the bulk work function from the preceding paragraph (which is equal to $-10.6$ meV/165 K $= -6.4 \cdot 10^{-5}$ eV/K). This reiterates the statement above that the smaller clusters' ionization energies, akin to their thermal expansion, have a stronger temperature response. While no experimental data on smaller clusters are available, molecular dynamics simulations on Al$_6$ and Al$_7$ [36] qualitatively suggest a continuation of this trend (a calculated $dI_{6,7}/dT$ slope of $\sim -20 \cdot 10^{-5}$ eV/K over the range of $T=0$ K - 800 K).

## IV. SUMMARY

We have measured accurate photoionization efficiency curves for a series of free size-resolved and temperature-controlled aluminum nanoclusters. Variation of ionization energies with size reflects the orbital electronic structure of these particles, while their variation with temperature provides insight into the influence of the ionic framework onto the electronic states (e.g., via thermal expansion). The efficacy of measurements employing free clusters is that they make it possible to detect and quantify such subtle effects.

The observations include the following:

- Dips in the ionization energies at specific cluster sizes provide a good match with "superatom" model shell closings. However, in agreement with other work, there is also evidence for shell splitting and/or mixing effected by the ionic core.

- Extrapolation of the nanocluster ionization energies to the bulk limit enabled us to resolve and accurately determine the decrease in the polycrystalline work function $\varphi$ as the metal temperature rises from the lower experimental limit of 65 K to the upper limit of 230 K. The -0.23% drop in the value of $\varphi$ is in excellent agreement with theoretical prediction based on the volume thermal expansion of aluminum.

- The average thermal shifts of the smaller clusters' ionization energies appear stronger, which is congruent with the fact that they are expected to experience greater thermal expansion.

In future work, it would be useful to explore and understand thermal shifts in ionization energies of nanoclusters of yet smaller sizes and consequently greater surface-to-volume ratios. It would also be interesting to investigate analogous effects in mixed clusters of carefully controlled composition. For example, alloying bulk aluminum with Be, Si or Cu can alter the thermal expansion coefficients by as much as a factor of two [37]. Therefore one may expect to detect parallel alterations in the thermal ionization energy shifts. Such a measurement could provide a means of quantifying cluster thermal expansion, which is presently experimentally inaccessible, as well as general insight into nanoscale alloying with atom-by-atom accuracy.

## ACKNOWLEDGEMENTS

We would like to thank C. Huang and Dr. A. Liang for their help with the experiment. This work was supported by the U.S. National Science Foundation under Grant No. DMR-1206334.

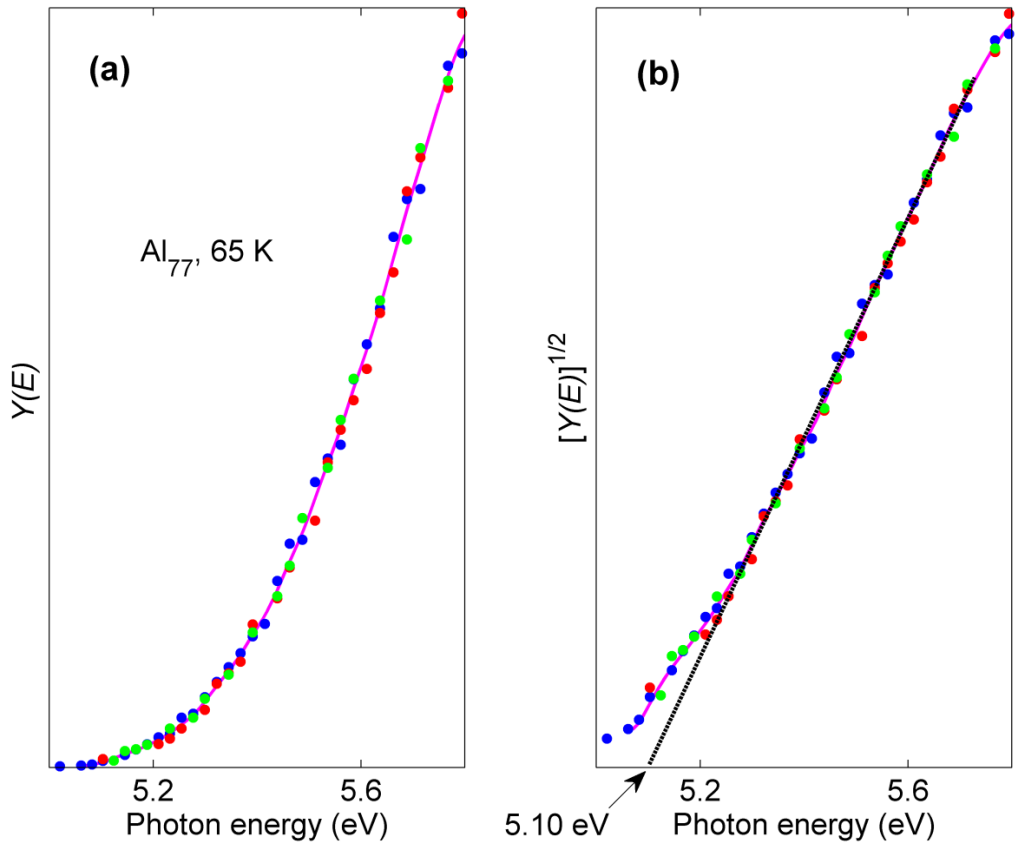

**Fig. 1.** (a) A sample photoionization yield curve, $Y_{77}(E)$, acquired at nanocluster temperature of 65 K. Different color dots correspond to data acquired in different runs. The red line demonstrates the match of Eq. (1) to the shape of the experimental curve. (b) Linear fit to the square root of the data in (a) showing the resulting threshold $I_{77}$.

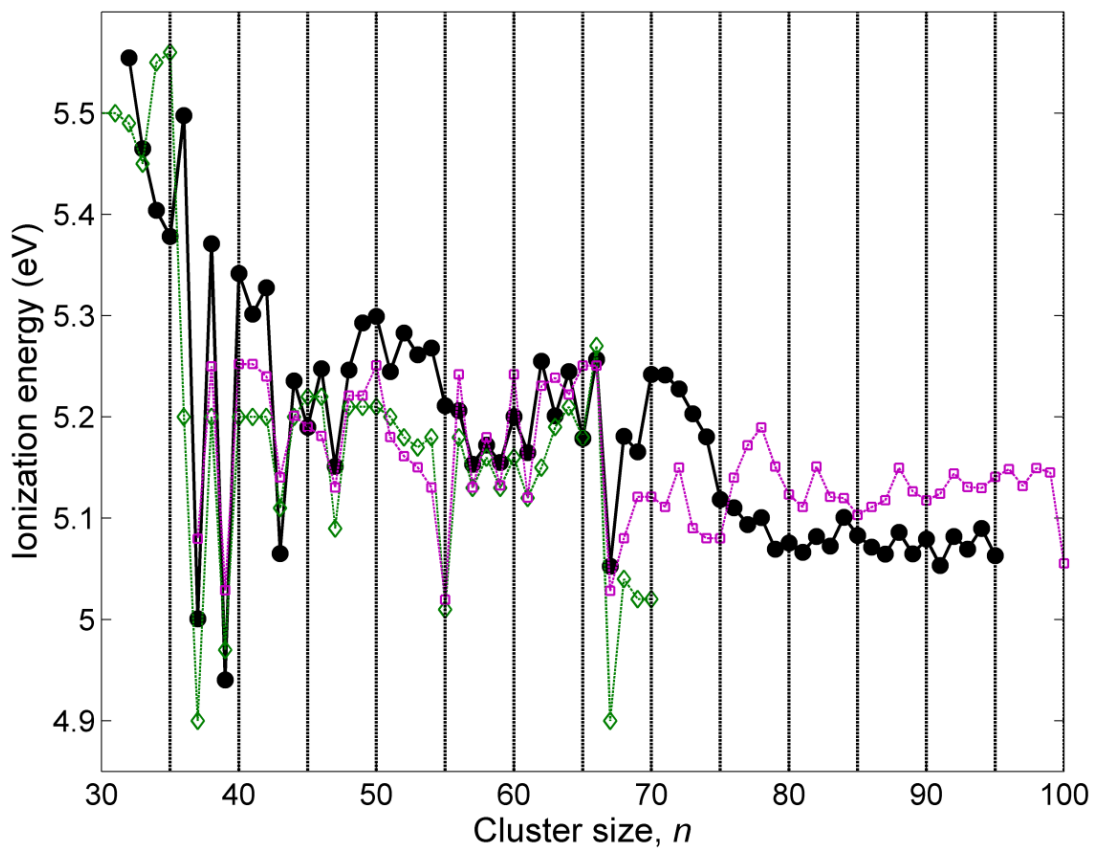

**Fig. 2.** Al$_n$ ionization energies. Solid dots: temperature-averaged values from the present experiment. Open diamonds: Refs. [27,28], open squares: Ref. [29]. Dips in $I_n$ reflect the nanoclusters' electronic shell pattern.

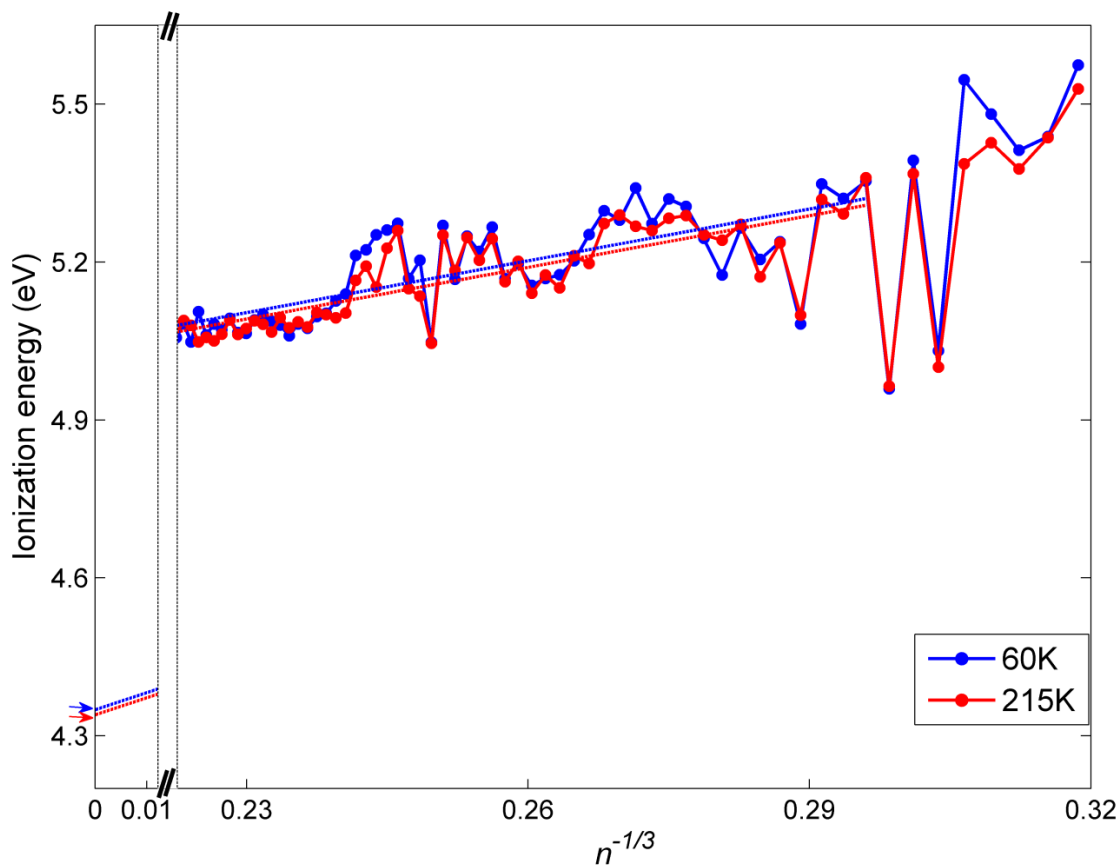

**Fig. 3.** Al$_n$ ionization energies measured for two different cluster temperatures. The dashed lines are linear fits to the two sets of data for $n \geq 40$ (the size estimated to correspond to the onset of metallic behavior of aluminum nanoclusters [1,7,8]). The lines follow the dependence shown in Eq. (2) with a metal-like slope of $\alpha=0.34$. The left-hand side of the plot shows the extrapolation of the data to the bulk limit, with the arrows marking the corresponding work functions.

# Nanocluster ionization energies and work function of aluminum, and their temperature dependence

Avik Halder and Vitaly V. Kresin

*Department of Physics and Astronomy, University of Southern California,*
*Los Angeles, California 90089-0484, USA*

**Fig. S1.** The individual thermal shift plots for each Al$_n$ nanocluster, $n$=32-95, are shown below. The blue dots are thresholds for cluster temperatures of 65K, 90K, 120K, 170K and 230K, and the red lines are linear fits to guide the eye.

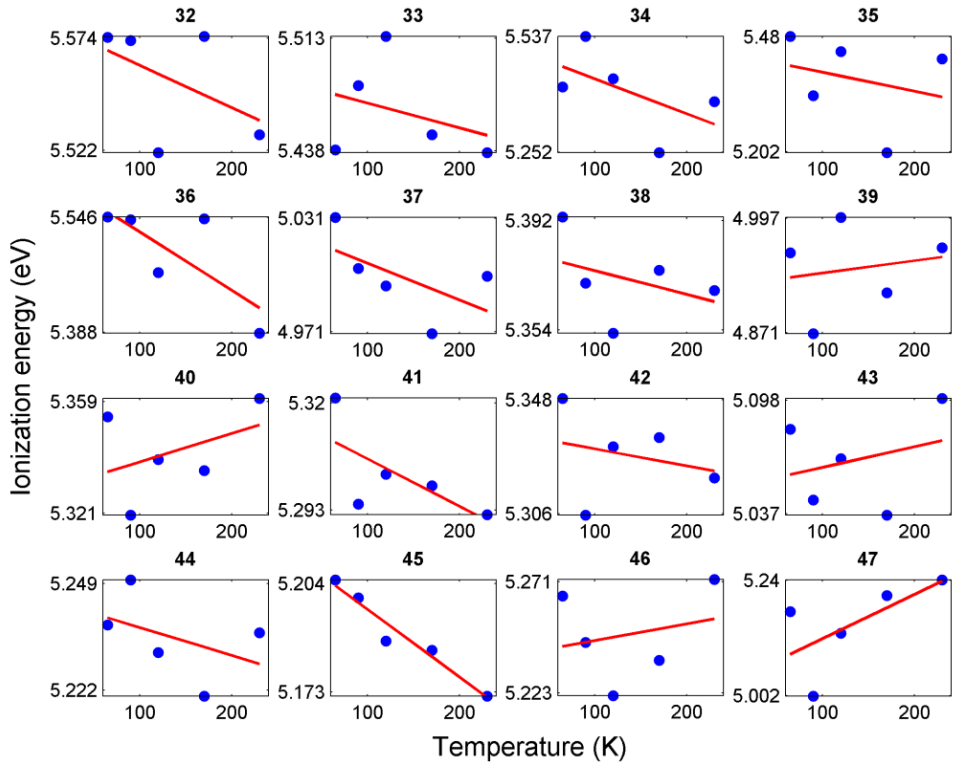

i

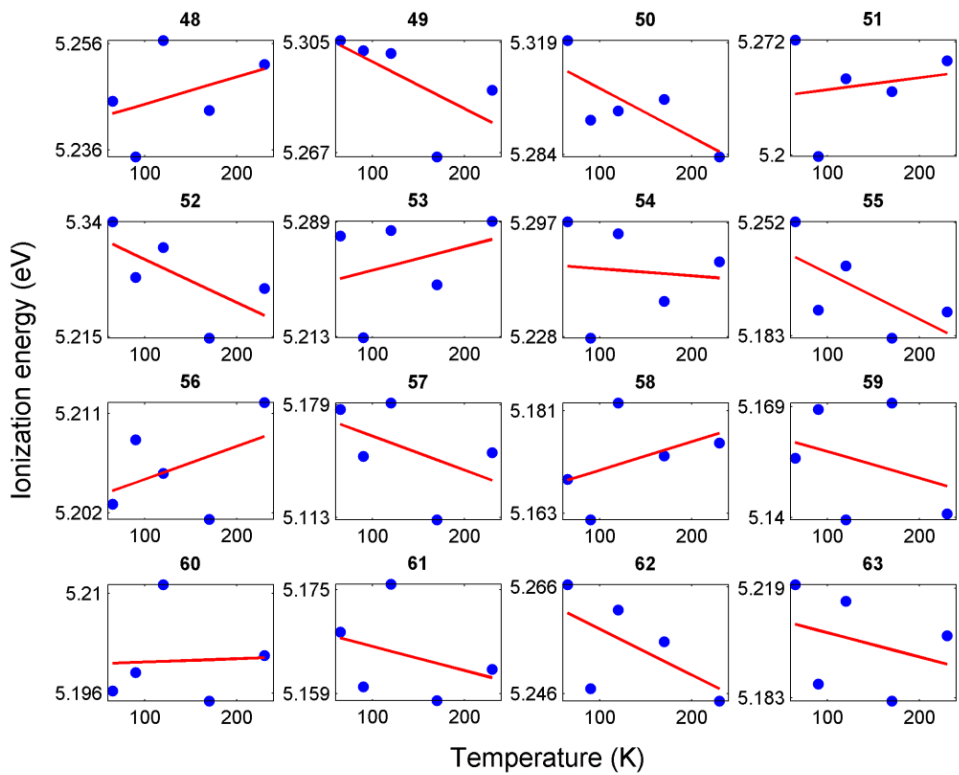

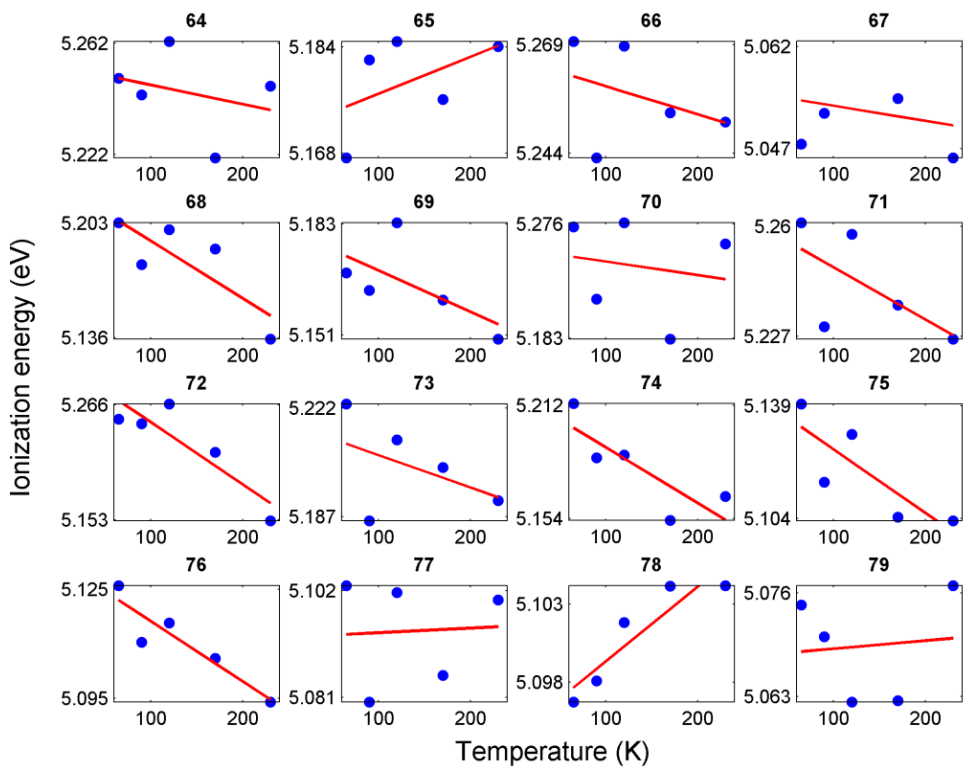

ii

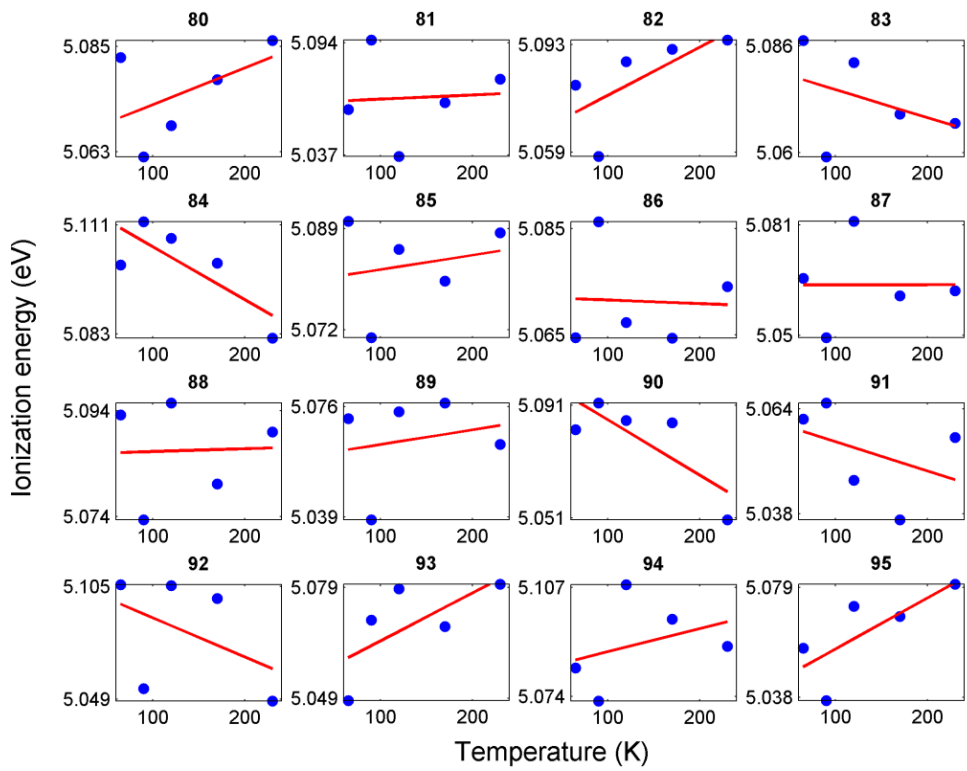

**Fig. S2.** Al$_n$ ionization energies for two different cluster temperatures (blue: 65 K, red: 230 K). The plot is analogous to Fig. 3 in the main text, but here the abscissa is linear in cluster size.

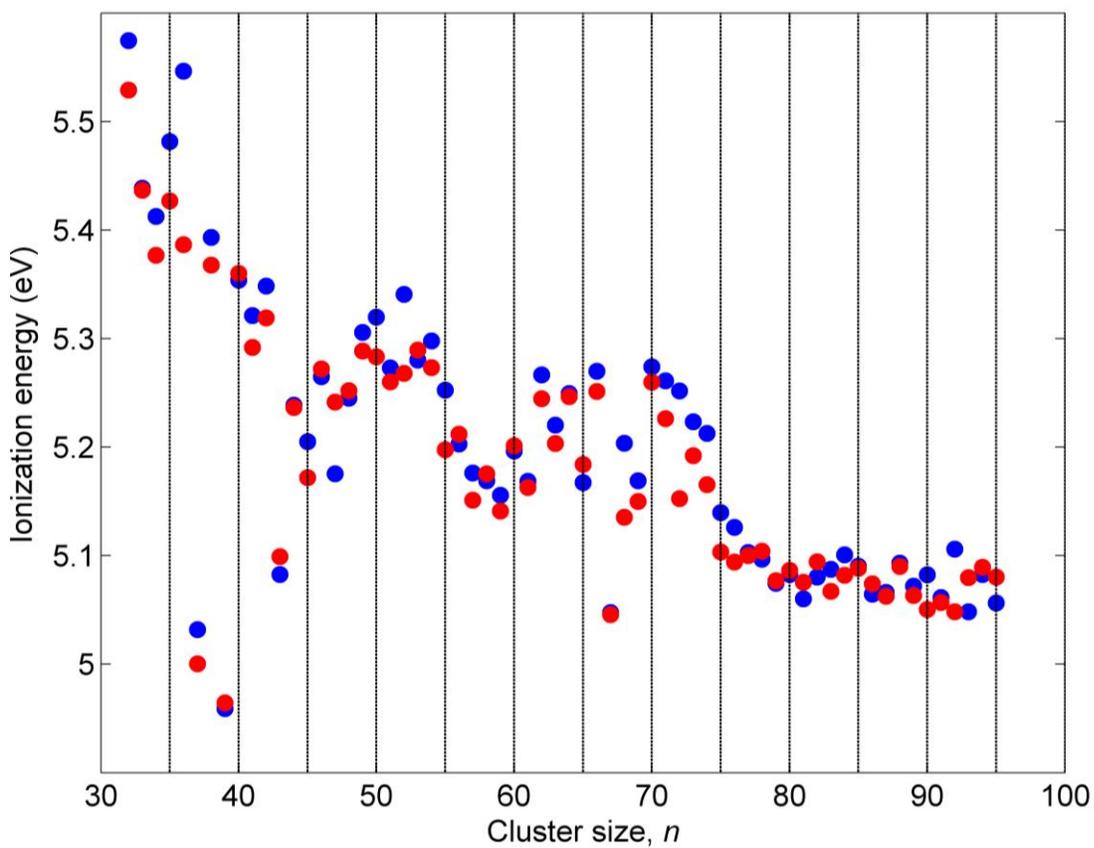